\documentclass[pss,fleqn]{w-art}
\usepackage{amsmath}
\usepackage{amssymb}
\usepackage{times}
\usepackage{w-thm}
\usepackage{graphicx}% Include figure files 
\usepackage{dcolumn}% Align table columns on decimal point
\usepackage{bm}% bold math
\usepackage{color}% perhaps for xfig exports needed
\usepackage{epsfig}%perhaps necessary for xfig exports
\usepackage{psfrag}
\usepackage{epstopdf}
\usepackage{hyperref}
\usepackage{cite}

\theoremstyle{plain}

\theoremstyle{definition}

\chardef\bslash=`\\ % p. 424, TeXbook

\begin{document}
\newcommand{\commute}[2]{\left[#1,#2\right]}
\newcommand{\bra}[1]{\left\langle #1\right|}
\newcommand{\ket}[1]{\left|#1\right\rangle }
\newcommand{\anticommute}[2]{\left\{  #1,#2\right\}  }
\renewcommand{\arraystretch}{2}

%%    The information for the title page will be placed between
%%    \begin{document} and \maketitle. The order of most entries
%%    is determined by the class file and can not be changed by
%%    rearranging them. The maketitle command follows after the
%%    abstract.
%%
%%    Most of the following commands will be completed by the publisher.
%%
%%    The copyrightyear is defined in the .clo file as the first argument
%%    of the copyrightinfo command. If the copyrightyear differs from that
%%    value it might be adjusted by the following definition:
%%
\renewcommand{\copyrightyear}{2006}% uncomment to change the copyrightyear.
%%
%\DOIsuffix{theDOIsuffix}
%%
%% issueinfo for header and copyright line
\Volume{XX}
\Issue{X}
\Month{XX}
\Year{2006}
%%
%%    First and last pagenumber of the article. If the option
%%    'autolastpage' is set (default) the second argument may be left empty.
\pagespan{1}{}
%%
%%    Dates will be filled in by the publisher. The 'reviseddate' and
%%    'dateposted' (Published online) entry may be left empty.
\Receiveddate{\today}
\Reviseddate{\today}
%\Accepteddate{\today}
%\Dateposted{}
%%
\keywords{Quantum dot, quantum computing, double dot, parity, Bell state, measurement, spin-orbit interaction, hyperfine interaction, decoherence, dephasing, relaxation, spin}

\subjclass[pacs]{73.21.La, 73.23.Hk, 71.70.Ej, 72.25.Rb, 71.70.Jp}

%% \pretitle{Editor's Choice}

%% We have a short and a long form for the title. The short form
%% (optional argument) goes into the running head.

\title[Measurement, control, and decay\ldots]{Measurement, control, and decay of quantum-dot spins}

%% Please do not enter footnotes or \inst{}-notes into the optional
%% argument of the author command. The optional argument will go into
%% the header.  If there is only one address the marker \inst{x} may be
%% omitted.

%% Information for the first author.
\author[W. A.  Coish]{W. A.  Coish\inst{1}} \address[\inst{1}]{University of Basel, department of Physics and Astronomy, Klingelbergstrasse 82, 4056 Basel}
%%
%%    Information for the second author
\author[V. N. Golovach]{Vitaly N. Golovach\inst{1}}
%%    Information for the third author
\author[J. C. Egues]{J. Carlos Egues\inst{1,2}}
\address[\inst{2}]{Departamento de F\'{\i}sica e Inform\'atica, Instituto de F\'{\i}sica de S\~ao Carlos, Universidade de S\~ao Paulo, 13560-970 S\~ao Carlos, S\~ao Paulo, Brazil}
%%    Information for the fifth author
\author[D. Loss]{Daniel Loss\inst{1}}
%%
%%    \dedicatory{This is a dedicatory.}
\begin{abstract}
In this review we discuss a recent proposal to perform partial Bell-state (parity) measurements on two-electron spin states for electrons confined to quantum dots.  The realization of this proposal would allow for a physical implementation of measurement-based quantum computing.  In addition, we consider the primary sources of energy relaxation and decoherence which provide the ultimate limit to all proposals for quantum information processing using electron spins in quantum dots.  We give an account of the Hamiltonians used for the most important interactions (spin-orbit and hyperfine) and survey some of the recent work done to understand dynamics, control, and decoherence under the action of these Hamiltonians. We conclude the review with a table of important decay times found in experiment, and relate these time scales to the potential viability of measurement-based quantum computing.       
\end{abstract}
%% maketitle must follow the abstract.
\maketitle                   % Produces the title.

%% If there is not enough space inside the running head
%% for all authors including the title you may provide
%% the leftmark in one of the following three forms:

%% \renewcommand{\leftmark}
%% {First Author: A Short Title}

%% \renewcommand{\leftmark}
%% {First Author and Second Author: A Short Title}

%% \renewcommand{\leftmark}
%% {First Author et al.: A Short Title}

%% \tableofcontents  % Produces the table of contents.
\section{Introduction}

By now there have been very many proposals to implement quantum information processing using semiconductor quantum dots. Some of these proposals include single electron spins confined to quantum dots, coupled by the exchange interaction \cite{loss:1998a,burkard:1999a}, quantum-dot spins coupled via optical cavity modes \cite{imamoglu:1999a}, ferroelectrically coupled Si:Ge quantum dots \cite{levy:2001a}, exchange-coupled Si quantum dots \cite{friesen:2003a}, excitons \cite{krenner:2005a} (in which coupled molecular states can now be generated and manipulated \cite{krenner:2006a}), coupled two-spin encoded qubits \cite{levy:2002a} (a topic which has received quite some recent attention \cite{petta:2005a,taylor:2005b,taylor:2005a,burkard:2006a,hanson:2006a}), many-spin cluster qubits composed of antiferromagnetically-coupled spin chains \cite{meier:2003a,meier:2003b} (some promising new experiments show that such chains can be controllably built atom-by-atom on a surface \cite{hirjibehedin:2006a}), and potentially the use of the heavy-hole \cite{bulaev:2005b} or nuclear \cite{taylor:2003a} spin as a carrier of long-lived quantum information.  One- and two-qubit operations are sufficient for universal quantum computation \cite{barenco:1995a}. Many of these proposals are therefore designed around the common lore that universal quantum computation \emph{requires} two-qubit interactions.  In less-conventional \emph{measurement}-based quantum computing, universal quantum computation can be achieved without two-qubit interactions.  In these schemes, two-qubit operations can still be implemented provided certain measurements can be performed.  Recently, the details have been worked out for a ``free electron'' quantum computing scheme \cite{beenakker:2004a}, which requires parity measurements to be performed on the number operator for pairs of fermionic modes.  In the first section of this review, we describe a possible physical implementation of related parity measurements for electron spins in quantum dots, as proposed recently in Ref. \cite{engel:2005a} (see also Ref. \cite{trauzettel:2006a} for a proposal to perform parity measurements on \emph{charge} qubits).

Decoherence remains an important aspect of the proposal of Ref. \cite{engel:2005a}.  As in conventional quantum computing, in measurement-based quantum computing, the qubit states must remain coherent during single-qubit operations.  We therefore devote the rest of the review to the two interactions that are the strongest source of spin decoherence in quantum dots: the spin-orbit and hyperfine interactions.

The rest of this review is organized as follows: In Sec. 2 we describe the physical implementation of parity measurements proposed in Ref. \cite{engel:2005a}.  In Sec. 3 we give a derivation of the spin-orbit Hamiltonians relevant for quantum dots and show how they give rise to energy relaxation and decoherence.  In Sec. 4 we survey some recent studies of quantum-dot electron spin dynamics in the presence of the hyperfine interaction.  In Sec. 5 we tabulate some of the most important decay times that have been measured for electrons in quantum dots, due to spin-orbit coupling, hyperfine coupling, and charge fluctuations.  We then describe how these coherence times apply to the proposal discussed in Sec. 2 and suggest ways to minimize unwanted decoherence.  

\section{Spin parity meter}
\label{sec:BellState}

Most proposed quantum computation schemes crucially rely on the possibility of judiciously controlling the interaction between qubits. A controlled qubit-qubit interaction is an important ingredient to the implementation of two-qubit logic gates (e.g., {\sc cnot}), which, together with single-qubit rotations constitutes a universal set of quantum logic gates (the fourth DiVincenzo criterion), sufficient to perform any quantum computation. In the Loss-DiVincenzo proposal for quantum computation with electron spins in quantum dots as qubits, the qubit-qubit interaction is conveniently provided by the exchange interaction between the electrons and can, in principle, be controlled via additional pulsed gate electrodes changing the potential barrier between two neighboring dots. The Loss-DiVincenzo quantum CPU satisfies all five relevant requirements (``DiVincenzo criteria'') for quantum computer architectures. However, the precise control of the gate electrode controlling the interaction between the spin qubits poses an additional technological challenge for the implementation of spin-based quantum computing. More specifically, the logic-gate operation time $\tau_s$ for both single-qubit rotation and two-qubit manipulations (e.g., {\sc swap}) must be much shorter than the decoherence time $T_2$ such that $\tau_s/T_2<\epsilon$ where $\epsilon=10^{-3}$-$10^{-4}$\cite{steane:2003a} is the maximum tolerable error rate for error-correction schemes to remain effective. For two-qubit operations, $\tau_s$ is given by the time the pulsed gate electrode is turned on.  In addition, any error in the pulse duration (and/or inter-qubit coupling strength) directly translates into errors in two-qubit operations. 

Beenakker et al. \cite{beenakker:2004a} have recently proposed a quantum computation scheme involving only non-interacting spin qubits. These authors use both full and partial Bell-state measurements on two spin-qubit states to construct a {\sc cnot} gate which, together with single spin qubit rotations, suffices for universal quantum computation. Interestingly, these authors point out that partial Bell-state measurements are sufficient for efficient (i.e., exponential speedup as compared to a classical computation)\footnote{We note that in Ref. \cite{beenakker:2004a} the authors overcome the no-go theorem\cite{terhal:2002a} that precludes efficient quantum computation with free fermions by considering both the spin and the charge degrees of freedom of the spin qubits, i.e., by performing both charge and spin measurements.} and deterministic (i.e., the logic gates operate with 100\% success) quantum computation. A partial Bell-state measurement is a projective measurement in which only the \emph{spin parity} (the total spin projection along the $z$-axis) of a two-spin qubit state is measured, but the individual spin components are still unknown. For instance, the singlet  and the entangled triplet   two-qubit states both have parity $p=0$, while the spin-parallel triplets have parity $p=1$. Hence, a spin-parity measurement should leave intact linear combinations of the Bell states which belong to the subspace $p=0$, as well as those of the Bell states for which $p=1$. We note that in quantum optics with qubits encoded in photon polarization states, Knill et al. \cite{knill:2001a} have earlier shown the feasibility of nearly deterministic efficient quantum computation with free photons.  The measurement-based quantum computation scheme of Ref. \cite{beenakker:2004a} completely obviates the need for qubit-qubit interaction, thus making the implementation of spin-based quantum computation less technologically challenging. Next we briefly discuss a recent proposal for a spin-parity meter using a realistic quantum dot setup \cite{engel:2005a}.
 
\begin{figure}[tbh]
\includegraphics[width=\textwidth]{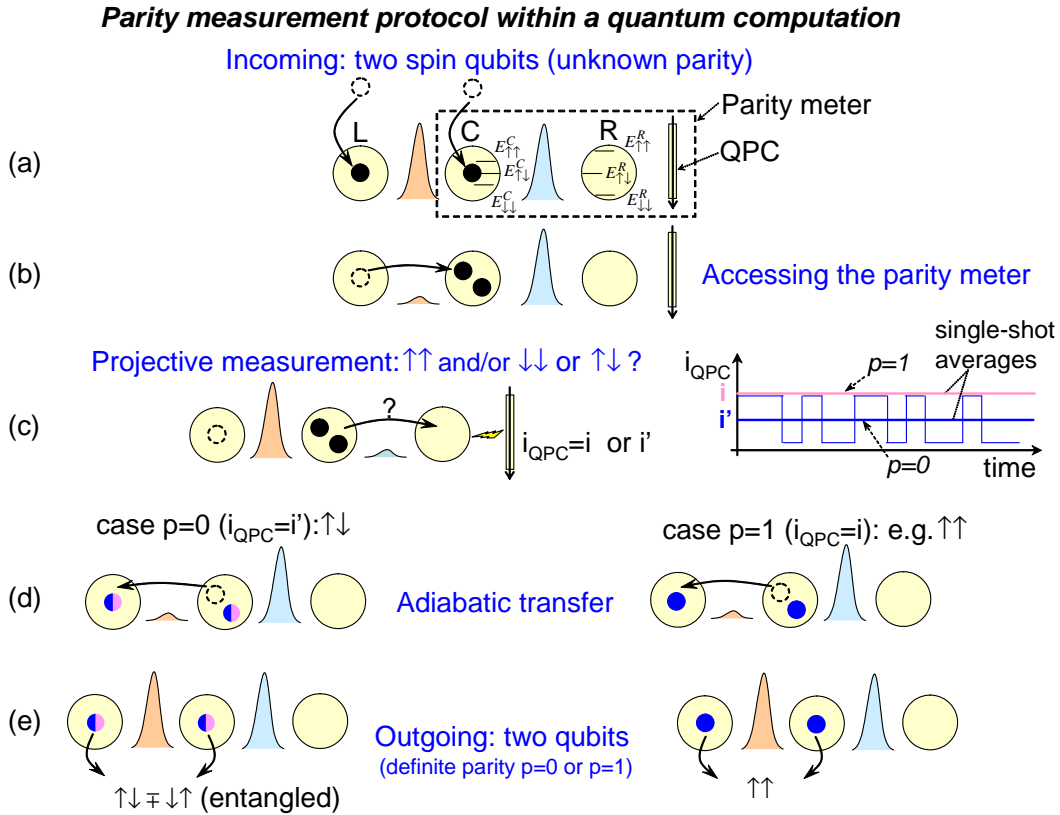}
\caption{Illustrative scheme integrating the spin-parity meter of Engel and Loss \cite{engel:2005a} into a quantum computation.  Note that the right dot here [part of the spin-parity meter] acts as an ancillary dot in the quantum processing.}
\label{fig:SpinParityMeasurement}
\end{figure}

The spin parity meter of Engel and Loss \cite{engel:2005a} -- realizable with existing state-of-the-art quantum-dot nanotechnology -- performs projective parity measurements on two incoming spin qubits and hence should enable free-qubit measurement-based quantum computation. Figure \ref{fig:SpinParityMeasurement}(a) schematically depicts two spin qubits of unknown parity entering the two left-most quantum dots (L, C) of a set of three quantum dots (L,C,R). These two ``flying qubits'' could be part of some ongoing quantum processing. The two rightmost dots (C, R) together with the neighboring quantum point contact QPC (all enclosed in a dashed rectangle) constitute the spin parity meter. All three dots are separated by potential barriers whose heights can be varied via gate electrodes (note that here we do not need to control the barrier heights with extreme precision so as to perform quantum gate operations directly -- only to the extent of enabling tunneling between the dots). The L and C dots are assumed to be identical (i.e., they have the same level structure). The C and R dots, on the other hand, are engineered to have \emph{different} Zeeman energies\footnote{Different Zeeman splittings on the two dots could be achieved, for example, through g-factor modulation\cite{salis:2001a}, the inclusion of magnetic layers\cite{myers:2005a}, with nearby ferromagnetic dots\cite{loss:1998a}, or by polarizing the nuclear spins in one dot to achieve a substantial Overhauser field (see Sec. 4, below)} for the parallel ($p=1$) spin pairs $E^{\uparrow\uparrow}_\mathrm{C}\ne E^{\downarrow\downarrow}_\mathrm{C}\ne E^{\uparrow\uparrow}_\mathrm{R}\ne E^{\downarrow\downarrow}_\mathrm{R}$ and equal Zeeman energies for the antiparallel ($p=0$) spin pairs $E^{\uparrow\downarrow}_\mathrm{C} = E^{\uparrow\downarrow}_\mathrm{R}$. In addition, degeneracy between the singlet $\ket{S}$ and the triplet $\ket{T_0}$ levels, both of which have energy $E^{\uparrow\downarrow}_\mathrm{C}$ in the central dot and in the right dot, can be attained via an applied out-of-plane magnetic field \cite{burkard:1999a,burkard:2000c,zumbuehl:2004a}.  The spin parity measurement proceeds as follows: after the two electrons of unknown parity are loaded into the (L,C) dots (a), they are adiabatically transferred into the parity meter by lowering the  potential barrier between the L and C dots (b) thus occupying the proper energy levels consistent with their symmetry (singlet or triplet).  Because of the particular level structure of the C and R dots, only electron pairs with parity $p=0$ (singlet and/or entangled triplet states) can tunnel into the R dot (since $E^{\uparrow\downarrow}_\mathrm{C}=E^{\uparrow\downarrow}_\mathrm{R}$) when the potential barrier between the C and the R dots is pulsed lower for the parity measurement to take place (c).  Hence, if the loaded pair has parity $p=0$, it will tunnel back and forth between the right dot R and center dot C, thus electrostatically modulating the current flowing in the QPC near it [see a sketch of the $i_\mathrm{QPC}$ signal in (c)], which would then register an average current $i_\mathrm{QPC} = i^\prime$. If the loaded pair has parity $p=1$ it does not tunnel into R at all due to the energy mismatch following from the distinct Zeeman energies and the readout in the QPC is $i_\mathrm{QPC} = i$. Hence, an \emph{average} charge current measurement via a QPC uniquely identifies the spin parity of an unknown two-spin qubit state. This measurement exemplifies '¡Èspin-to-charge'¡É conversion in which the spin information is extracted via  charge detection using a QPC. Note that this is a single-shot measurement, that is, it requires only one loaded pair at a time for each average current measurement. After the parity measurement, the electron pair is adiabatically transferred\footnote{This transfer should be done non-adiabatically with respect to the Zeeman splitting ($E^{\uparrow\uparrow}-E^{\downarrow\downarrow}$) to preserve the same coherent superposition $a\ket{\uparrow\uparrow}+b\ket{\downarrow\downarrow}$ that was obtained after measurement, but adiabatically with respect to the orbital level spacing $\hbar\omega_0$ (see, for example, the discussion at the end of Sec. 3.1)} into the leftmost dots (L,C) so that some quantum calculation being carried out can be completed (d). Note that if $p=0$ the outgoing two-qubit state has total spin $z$-projection $|S_z|=0$ [(e), left panel] while it has spin $z$-projection $|S_z|=1$ if $p=1$ [(e), right panel]. 

As discussed in Ref. \cite{engel:2005a}, there are many important issues concerning decoherence (orbital and spin) as well as spin and charge relaxation for the operation of the spin parity meter described above.  Here we simply note that the spin parity meter is supposed to operate in a regime in which the measurement time (i.e., the time it takes for the QPC to provide a meaningful average current result) should be larger than the orbital dephasing time of the relevant right-left charge states, but shorter than the spin-decay time and charge relaxation time due to inelastic processes.  See also the discussion following Table \ref{tab:spindecaytimes} in Sec. \ref{sec:Conclusions}.  In the following two sections, we review the strongest sources of decoherence for quantum dots: the spin-orbit and hyperfine interactions.   

\section{Spin-orbit interaction in quantum dots}
\label{sec:SpinOrbit}

On the way to utilizing the electron spin for quantum information
purposes,
two major approaches are undertaken in systems with spin-orbit
interaction.
In one approach, the true electron spin is taken as the carrier
of information and the orbital motion of the electron generates an
internal
magnetic field due to the spin-orbit interaction.
Precise knowledge of the spin-orbit interaction and of the electron motion
allows one to predict the evolution of the electron spin in time.
This approach is efficient when the spin precession due to the
internal spin-orbit field is slow, so that preparation and measurement
of electrons of a given spin is possible.

In a different approach, the time-reversal symmetry of the electron
in the presence of spin-orbit interaction is exploited.
In the absence of external magnetic fields, the states of the electron are
doubly degenerate.
In spite of the fact that the spin is not conserved in the presence
of spin-orbit interaction, the time-reversal symmetry (Kramers theorem) 
allows one to encode the qubit into long-lived states.
When a Zeeman field is turned on, these states behave like a
usual spin-$1/2$ in the absence of spin-orbit interaction.
The preparation and measurement of Kramers doublets should be
performed with certain adiabaticity constraints which we discuss below.

\subsection{Spin-orbit interaction and time-reversal symmetry}
\label{sec:TRSymmetry}

The spin-orbit interaction in semiconductors originates from the
crystalline potential, which is a strongly oscillating function
of the electron position. The periodic electric field ``seen'' by 
the electron as it moves across the crystal lattice gives rise to 
an internal magnetic field, which in its turn couples to the
electron spin via the magnetic moment.
In general terms, the spin-orbit interaction (in gaussian units) reads
\begin{equation}
H_{SO}=\frac{g_s\mu_B}{4 m_0 c^2}
\left[\mbox{\boldmath $\nabla$} V(\mbox{\boldmath $r$})\times
\mbox{\boldmath $p$}\right]\cdot
\mbox{\boldmath $\sigma$}.
\label{HSOrel}
\end{equation}
Here, $g_s\approx 2$ is the free-electron g-factor,
$\mu_B$ is the Bohr magneton, $m_0$ is the bare electron mass, 
$c$ is the speed of light in vacuum,
$V(\mbox{\boldmath $r$})$ is the crystal (or any other) potential,
$\mbox{\boldmath $r$}$ and $\mbox{\boldmath $p$}$ are the electron 
coordinate and kinetic momentum.
The spin-orbit interaction has a relativistic nature ($\sim 1/c^2$) 
and it leads to small effects for electrons in vacuum.
In solid-state systems, however, the band nature of the electron plays an
important role.
In particular, the crystal potential leads to large
potential gradients $\mbox{\boldmath $\nabla$} V$, 
which results in a sizable spin-orbit interaction for the band electron.
It is important to note, however, that if the crystal potential $V({\bm
r})$
has inversion symmetry, then the spin-orbit interaction
averages to zero in each unit cell and results in no 
spin-splitting of electronic bands. 

In semiconductors with the zinc-blende crystal structure,
the spin-orbit interaction induces a spin-splitting
of the conduction band proportional to the
cube of the electron wave-vector ${\bm k}$
(Dresselhaus spin-orbit interaction\cite{dresselhaus:1955a}) 
\begin{equation}
H_{SO}=\frac{\alpha_c}{2\sqrt{2m_e^3E_{\rm
g}}}{\bm\varkappa}\cdot\mbox{\boldmath $\sigma$},
\label{Dresselhaus3D}
\end{equation}
where $\varkappa_x=\hbar^3 k_x\left(k_y^2-k_z^2\right)$ (other components
are obtained by cyclic permutations of indices),
$m_e$ is the electron effective mass, and $E_{\rm g}$ is the band gap.
The coupling constant $\alpha_c$ ($\approx 0.07\;\mbox{for
GaAs}$\cite{dyakonov:1986a,pikus:1984a}) 
arises due to the absence of inversion symmetry in the unit cell.
The Dresselhaus spin-orbit interaction is highly anisotropic.
The coordinate frame in Eq.~(\ref{Dresselhaus3D}) is chosen such that
the axes $x$, $y$, and $z$ are parallel to the main crystallographic
directions.
The spin-precession in the spin-orbit field (\ref{Dresselhaus3D})
can be accurately predicted provided the electron wave-vector ${\bm k}$ 
is conserved.
If the momentum scattering time $\tau_p$ is finite,
spin relaxation is controlled by the parameter
$\Omega\tau_p$, where $\Omega$ is a characteristic spin-precession
frequency.
For $\Omega\tau_p\gg 1$, the spin lifetime $\tau_\mathrm{spin}$ is on the order of
$\tau_p$,
whereas for $\Omega\tau_p\ll 1$, one has $1/\tau_\mathrm{spin}\sim\tau_p\Omega^2$ due
to
motional narrowing\cite{pikus:1984a}.

In lower-dimensional systems, the spin-orbit interaction
is obtained to lowest order by integrating out the electron 
motion along the confinement direction.
For example, in 2D, the Dresselhaus spin-orbit interaction
is given by Eq.~(\ref{Dresselhaus3D}) with ${\bm \varkappa}$ 
changed to\cite{dyakonov:1986a}
\begin{equation}
\varkappa_x=\frac{\hbar^3}{d^2}
\left(2n_x(n_yk_y-n_zk_z)+(n_y^2-n_z^2)k_x\right),
\end{equation}
where $k_xn_x+k_yn_y+k_zn_z=0$, ${\bm n}$ is the normal vector to the 2D
plane, and
$d$ is a characteristic size of the electron wave function 
along the quantization direction, {\em i.e.}
$1/d^2=\langle({\bm k}\cdot{\bm n})^2\rangle$.
For the most commonly used GaAs heterostructures grown
along $[001]$, the Dresselhaus interaction takes the
form
\begin{equation}
H_{SO}^D=\hbar\beta\left(-k_x\sigma_x+k_y\sigma_y\right),
\label{HSOD}
\end{equation}
where $\beta=\hbar^2\alpha_c/2d^2\sqrt{2m_e^3E_{\rm g}}$.
In heterostructures, an additional spin-splitting 
proportional to ${\bm k}$ 
(Rashba spin-orbit interaction\cite{bychkov:1984a})
arises due to the lack of inversion symmetry at the
interface between two materials.
By symmetry, one can obtain the Rashba
spin-orbit interaction from Eq.~(\ref{HSOrel})
assuming that $-\mbox{\boldmath $\nabla$} V$
is a homogeneous electric field along the
quantization direction,
\begin{equation}
H_{SO}^R=\hbar\alpha\left(k_x\sigma_y-k_y\sigma_x\right).
\label{HSOR}
\end{equation}
However, a microscopic derivation of Eq.~(\ref{HSOR}) shows
that $\alpha$ is dominated by the offsets of valence bands
in GaAs/GaAlAs heterostructures\cite{zawadzki:2004a}.
The Rashba spin-orbit interaction is attractive for
applications in spintronics, because the coupling constant 
$\alpha$ can, in principle, be electrically controlled.
A gate (or carrier concentration) control of $\alpha$ 
has been demonstrated in InGaAs heterostructures\cite{nitta:1997a}.

In quantum dots defined at the interface between two 
materials, the Dresselhaus (\ref{HSOD}) and Rashba 
(\ref{HSOR}) spin-orbit interactions govern most of 
the spin-orbit-related effects.
Quite remarkably, the length-scales
associated with the spin-orbit interaction are independent of 
the electron orbital state in the quantum dot, 
because the spin-orbit interaction is linear in ${\bm k}$.
Introducing a tensor of inverse spin-orbit lengths\cite{golovach:2006a}
\begin{equation}
\left(\lambda_{SO}^{-1}\right)_{ij}=
\frac{m_e}{\hbar^2}\langle\psi_n|\frac{\partial^2H_{SO}}{\partial\sigma_i\partial
k_j}|\psi_n\rangle,
\end{equation}
one finds for $H_{SO}=H_{SO}^D+H_{SO}^R$ no dependence on
$|\psi_n\rangle$ [where above, $i,j=\left(x,y\right)$].
In this case, the wave function of a point-like electron (measured at position $\mathbf{r}_0$) has a simple
form
\begin{equation}
\psi_{{\bm r}_0\sigma}({\bm r},s)=
\mathcal{P}e^{-i\int{\bm \sigma}\cdot\lambda_{SO}^{-1}\cdot d{\bm r}_0}
\delta({\bm r}-{\bm r}_0)\chi_{\sigma}(s),
\label{psipath}
\end{equation}
where $\chi_{\sigma}(s)$ is a spinor and the 
exponent is ordered along the path of integration by the path ordering operator $\mathcal{P}$\footnote{$\mathcal{P}$ is analogous to the time ordering operator $\mathcal{T}$ that appears in the unitary time evolution: $U(t)=\mathcal{T}\exp\left[\int_0^t H(t^\prime)dt^\prime\right]$}.
The exponential factor in Eq.~(\ref{psipath}) 
is a spin-dependent phase and, of course, can be gauged away
for a fixed path.
The physical interpretation of this exponential factor
is as follows:
the electron spin is ``rotated'', due to the spin-orbit interaction, 
as it moves along a given path.
Since the spin-orbit interaction is time-reversal invariant,
going backwards along the same path undoes the spin rotation.
In a quantum dot, the size of the electron wave function is finite
and Eq.~(\ref{psipath}) holds only approximately.
Nevertheless, if the area of the dot is small compared
to $1/\det(\lambda_{SO}^{-1})$, then the path ordering in
Eq.~(\ref{psipath}) can be
ignored and one can write
\begin{equation}
\psi_{n\sigma}({\bm r},s)=e^{-i{\bm \sigma}\cdot\lambda_{SO}^{-1}\cdot
{\bm r}}
\psi_n({\bm r})\chi_{\sigma}(s),
\label{psinopath}
\end{equation}
where $n$ is a set of orbital quantum numbers.
Equation~(\ref{psinopath}) has been widely used in the
literature~\cite{khaetskii:2000a,halperin:2001a,aleiner:2001a}
to gauge away the spin-orbit interaction to leading order.

To obtain a more rigorous treatment of the spin-orbit interaction in
quantum dots, one can attempt to diagonalize the quantum dot
Hamiltonian using perturbation theory.
In this case, the electron wave function is presented as follows
\begin{equation}
\psi_{n\sigma}({\bm r},s)=e^{-S}\psi_n({\bm r})\chi_{\sigma}(s),
\end{equation}
where $S$ is an operator that acts on both ${\bm r}$ and $\sigma$.
Expanding $S$ in terms of $H_{SO}$ one obtains at leading order
$[H_0,S]=H_{SO}$, where $H_0$ is the quantum dot Hamiltonian
without the spin-orbit interaction.
The transformation matrix $S$ can be evaluated to any desired order
using standard methods of perturbation theory.
According to the Kramers theorem, the energy levels of an
electron in a quantum dot are doubly degenerate,
\begin{equation}
E_{n\uparrow}=E_{\bar n\downarrow},
\end{equation}
where $\bar n$ denotes a state conjugated to $n$.
For the ground state, $n\equiv \bar n$ and 
the Kramers doublet is closely related to a spin doublet.

For quantum computing with electron spins in quantum dots,
it is convenient to work with Kramers doublets rather than
with the true electron spin.
The reason for this is a large energy scale responsible
for ``dressing'' the true spin with orbital degrees
of freedom by the spin-orbit interaction.
This energy scale is on the order of the size quantization
energy, $\hbar\omega_0$, which is considerably larger than
any energy scale used in quantum computing with spins.
Using Kramers doublets to encode the qubit, therefore, also requires
that the manipulation of the qubit be performed adiabatically
with respect to $\hbar\omega_0$, to prevent excitations to
higher-energy orbital levels.

\subsection{Spin relaxation}

The spin-orbit interaction considered in Sec.~\ref{sec:TRSymmetry}
allows coupling of the electron spin to any environmental
degree of freedom that couples to the electron charge.
During a spin-flip, the angular momentum 
associated with the change of spin is given to the lattice, 
while the spin-flip energy is dissipated in the environment.
For single spins in quantum dots, a large variety of 
environments have been considered, such as  
phonons\cite{khaetskii:2001a,golovach:2004a,erlingsson:2001a,sanjose:2006a}, 
particle-hole excitations at the Fermi level\cite{borhani:2006a,levitov:2003a}, 
a functioning QPC nearby the quantum dot\cite{borhani:2006a}, charge fluctuations giving rise to Nyquist noise\cite{marquardt:2005a,sanjose:2006a}, etc. Additionally, there are spin-flip processes in which the energy is carried away by phonons, but the angular momentum is given to nuclear spins in the semiconductor \cite{erlingsson:2001a}, or where both energy and momentum are given to the nuclear spin system \cite{khaetskii:2002a,coish:2004a} (see Sec. \ref{sec:Hyperfine}, below).

Measurements of the spin-relaxation time\cite{elzerman:2004a,kroutvar:2004a} are in
good agreement with the phonon-induced spin relaxation\cite{khaetskii:2001a,golovach:2004a},
mediated by a combination of the spin-orbit and Zeeman interactions.
A characteristic feature of the mechanism in Refs.~\cite{khaetskii:2001a,golovach:2004a}
is a $B^5$-dependence of the spin relaxation rate on the magnetic field.
Such a dependence is observed in both gated GaAs quantum 
dots\cite{amasha:2006a}
and self-assembled InGaAs quantum dots\cite{kroutvar:2004a}.

At zero magnetic field, the lowest-energy Kramers doublet
in a quantum dot is well protected from decoherence by charge
environments.
By their nature, most environments (except $1/f$-noise and nuclear spins)
have vanishing spectral functions at low frequencies and temperatures.
As a result, a first-order process of energy exchange with the environment
becomes inefficient at low magnetic fields and the second-order processes
dominate.
A second order process requires available excitations in the environment.
Most environments, however, tend to ``freeze out'' at low temperatures.
Furthermore, the Kramers doublets
decouple from charge environments at first order in $H_{SO}$
and in the limit of zero magnetic field.
To illustrate this, we consider below a toy model for decoherence, based
on Eq.~(\ref{psipath}).

Neglecting corrections due to finite quantum dot size, we write down the
Kramers doublet
as follows
\begin{equation}
\psi_{\sigma}({\bm r},s)=
\mathcal{P}e^{-i\int{\bm \sigma}\cdot\lambda_{SO}^{-1}\cdot d{\bm r}_0(t)}
\psi\left({\bm r}-{\bm r}_0(t)\right)\chi_{\sigma}(s),
\label{psipath2}
\end{equation}
where ${\bm r}_0(t)$ is a fluctuating (quantum) field due to
the charge environment, and $\psi({\bm r})$ is the ground orbital state.
In the absence of decoherence, we choose ${\bm r}_0=0$ and
set the exponential factor to unity.
Equation (\ref{psipath2}) describes the evolution of a Kramers doublet
in the presence of a charge environment that couples to the center
of the electron charge distribution.
${\bm r}_0(t)$ fluctuates around its zero mean value and induces
a random spin rotation due to the exponential factor (ordered
along the path of $d{\bm r}_0$).

Next we assume that the fluctuations are small, ${\bm
r}_0\ll\lambda_{SO}$.
At first order, we can omit the ordering and obtain
\begin{equation}
\mathcal{P}e^{-i\int{\bm \sigma}\cdot\lambda_{SO}^{-1}\cdot d{\bm r}_0(t)}\approx
e^{-i{\bm \sigma}\cdot\lambda_{SO}^{-1}\cdot{\bm r}_0(t)}.
\end{equation}
To this approximation, the qubit undergoes reversible rotations, 
which on average maintain the qubit state coherence indefinitely.
Next we leave out the first-order terms since they do not lead to
decoherence.

To calculate the second-order terms,
we first find the infinitesimal change
of the exponential factor when increasing
the limit of integration by $\delta {\bm r}_0(t)$.
For this, we start with the following identity
\begin{equation}
\mathcal{P}e^{-i\int^{{\bm r}_0+\delta{\bm r}_0}{\bm
\sigma}\cdot\lambda_{SO}^{-1}\cdot d{\bm r}'_0}=
e^{-i{\bm \sigma}\cdot\lambda_{SO}^{-1}\cdot \delta{\bm r}_0}
\mathcal{P}e^{-i\int^{{\bm r}_0}{\bm \sigma}\cdot\lambda_{SO}^{-1}\cdot d{\bm r}'_0},
\end{equation}
where we expand each of the exponents on the right-hand-side in
power series. 
After dropping the irrelevant first-order terms, we obtain for the
infinitesimal change
\begin{equation}
-\left({\bm \sigma}\cdot\lambda_{SO}^{-1}\cdot\delta{\bm r}_0(t)\right)
\left({\bm \sigma}\cdot\lambda_{SO}^{-1}\cdot{\bm r}_0(t)\right).
\label{changebitinf}
\end{equation}
Since $\left(\lambda_{SO}^{-1}\right)_{ij}$ is two-dimensional, only the
Pauli matrix
$\sigma_z$ arises in Eq.~(\ref{changebitinf}) from products of $\sigma_x$
and $\sigma_y$.
Integrating over $\delta {\bm r}_0(t)$, we arrive at
\begin{equation}
\mathcal{P}e^{-i\int{\bm \sigma}\cdot\lambda_{SO}^{-1}\cdot d{\bm r}_0(t)}\approx
\mathcal{T}e^{i\sigma_z\det\left(\lambda_{SO}^{-1}\right)\int dt \left[{\bm
r}_0(t)\times \dot{\bm r}_0(t)\right]_z},
\label{finalexpsigmaz}
\end{equation}
where the subscript on $\left[\mathbf{v}\right]_z$ indicates the $z$-component of vector $\mathbf{v}$ and $\det\left(\lambda_{SO}^{-1}\right)=m_e^2(\alpha^2-\beta^2)/\hbar^2$
(up to a sign) is the inverse area spanned by the spin-orbit lengths 
$\lambda_\pm=\hbar/m_e\left(\beta\pm\alpha\right)$.
The physical meaning of Eq.~(\ref{finalexpsigmaz}) is as follows.
As the environment moves the dot center in the plane,
the electron spin acquires a phase proportional to the area
swept out by the radius-vector ${\bm r}_0(t)$.
Exploring the geometrical meaning of this result one can
further draw analogies with the Aharonov-Bohm flux
and the Berry phase~\cite{aleiner:2001a,sanjose:2006a}.
In Ref.~\cite{sanjose:2006a},
a detailed study of spin decoherence by different charge
environments and the second order spin-orbit interaction
is given.

In the presence of a magnetic field, the combination
of the Larmor precession and the fluctuating field 
${\bm r}_0(t)$ gives rise to spin relaxation at 
first order in $H_{SO}$.
A detailed analysis of this mechanism for both
relaxation and decoherence times is given in Ref.~\cite{golovach:2004a}.
Here, we rederive the same results in the
spirit of Eq.~(\ref{psipath}).

At finite magnetic field, the Kramers doublet
is split due to the Zeeman interaction and therefore
Eq.~(\ref{psipath2}) acquires a time-dependent
phase,
\begin{equation}
\psi_{\sigma}({\bm r},s,t)=
\mathcal{P}e^{-i\int{\bm \sigma}\cdot\lambda_{SO}^{-1}\cdot d{\bm r}_0(t)}
e^{-(i/2){\bm \sigma}\cdot{\bm \omega}_Lt}
\psi\left({\bm r}-{\bm r}_0(t)\right)\chi_{\sigma}(s),
\label{psipath3}
\end{equation}
where ${\bm \omega}_L$ is the Larmor frequency.
Next we would like to present Eq.~(\ref{psipath3})
in the following form
\begin{equation}
\psi_{\sigma}({\bm r},s,t)=
e^{-(i/2){\bm \sigma}\cdot{\bm \omega}_Lt}
\Phi_{\sigma}({\bm r},s)
\label{psi33}
\end{equation}
where $\Phi_{\sigma}({\bm r},s)$ changes slowly in time,
only due to spin decay.
The physical meaning of Eq.~(\ref{psi33}) 
is that of going to a rotating frame.
Further, the evolution of $\Phi_{\sigma}({\bm r},s)$ 
is presented in terms of the state in Eq.~(\ref{psipath2}),
since that state does not decay at first order in 
$H_{SO}$. 
As a result, we arrive at
\begin{equation}
\Phi_{\sigma}({\bm r},s)=U(t)
\psi_{\sigma}({\bm r},s),
\label{psi44}
\end{equation}
where $U(t)$ is the evolution operator that governs the decay.
At first order in ${\bm r}_0/\lambda_{SO}\ll 1$ we find
\begin{equation}
U(t)=\mathcal{T}e^{i\int dt\left[{\bm \omega}_L\times{\bm \Omega}(t)\right]\cdot{\bm
\sigma}},
\label{Uofteqnaa}
\end{equation}
where ${\Omega}_i(t)=\sum_j\left(\lambda_{SO}^{-1}\right)_{ij}r_{0j}(t)$.
Associating a Hamiltonian with the evolution operator in
Eq.~(\ref{Uofteqnaa}),
$H_{\rm int}=-\hbar[{\bm \omega}_L\times\Omega(t)]\cdot{\bm \sigma}$,
we arrive at an important result: the fluctuating fields generated by the
charge baths are transverse to the spin quantization axis defined by
${\bm \omega}_L$.
As a consequence, the dephasing processes, which are mediated by the
longitudinal
fluctuations, are deferred to higher orders of perturbation theory in
powers of
$H_{SO}$. If all other dephasing mechanisms could be arbitrarily suppressed, this would mean that the transverse-spin decay time is comparable to the longitudinal spin decay time (i.e. $T_2 = 2T_1$ \cite{golovach:2004a}) to leading order in the spin-orbit coupling.
In practice, the higher-order contributions are extremely weak and
different
dephasing mechanisms, such as the hyperfine interaction with the nuclear
spins dominate the spin decoherence rates in GaAs quantum dots.

\section{Hyperfine interaction in quantum dots}
\label{sec:Hyperfine}

Unlike the spin-orbit interaction, the hyperfine coupling of an electron spin to surrounding nuclear spins in a quantum dot can result in pure dephasing.  This pure dephasing is what limits the transverse spin coherence time in most quantum dots, in spite of the fact that the spin-flip ($T_1$) time is limited by the spin-orbit interaction and phonons at high magnetic fields (see Table \ref{tab:spindecaytimes} and the following discussion in Sec. \ref{sec:Conclusions}).  All III-V semiconducting materials have naturally occurring isotopes with nonzero nuclear spin \cite{schliemann:2003a}.  The hyperfine coupling is therefore ubiquitous in semiconductor quantum dots.  To make a quantum dot free of hyperfine-induced decoherence may require significant advances in materials-fabrication techniques to create coupled dots of isotopically purified material.  Gated double quantum dots for this purpose could be made in a two-dimensional electron gas (2DEG) embedded in a SiGe heterostructure \cite{friesen:2003a,slinker:2005a} or in carbon nanotubes \cite{mason:2004a,sapmaz:2006a,graeber:2006a}.  All of these elements ($\mathrm{Si},\mathrm{Ge},\mathrm{C}$) have naturally occurring isotopes with nuclear spin $I=0$ and isotopically purified samples of these materials can be obtained commercially \cite{cardona:2005a}.  Isotopic purification has been shown to increase spin coherence times for electrons bound to phosphorus donors in silicon \cite{abe:2004a,abe:2005a}.    

Traditional interaction-based quantum computing requires one-qubit and two-qubit operations, implemented by turning some interaction on and off.  In this context, it is therefore convenient to divide the decoherence of quantum-dot spins into single-spin and two-spin decoherence.  

\subsection{Single-spin decoherence}
\label{sec:HFOneSpin}

The Hamiltonian for a single electron in the lowest ($s$-type) orbital state of a quantum dot, interacting with surrounding nuclear spins via the Fermi contact hyperfine interaction is 
\begin{equation}
H_\mathrm{hf}=\mathbf{S}\cdot\left(g\mu_\mathrm{B}\mathbf{B}+\mathbf{h}\right);\,\,\,\,\,\mathbf{h}=\sum_i A_i\mathbf{I}_i, \label{eq:HFHamiltonian}
\end{equation}
where $\mathbf{S}$ is the spin-1/2 operator for the confined electron spin, $\mathbf{B}$ is an applied magnetic field, resulting in an electron-spin Zeeman splitting $g\mu_\mathrm{B}B$ ($|\mathbf{B}|=B$). $\mathbf{h}$ is the quantum nuclear field (``Overhauser operator") for the bath of nuclear spins, written in terms of nuclear spin operators $\mathbf{I}_i$ and the coupling constants $A_i$.  For an electron confined to a quantum dot in an orbital state with envelope wave function $\psi(\mathbf{r})$, the coupling constant to the nuclear spin at position $\mathbf{r}_i$ is $A_i=vA\left|\psi(\mathbf{r}_i)\right|^2$. Here, $v$ is the volume of a crystal unit cell containing one nuclear spin and $A\approx 90\,\mu eV$ gives the average hyperfine coupling constant in GaAs, weighted by the natural abundances of the three isotopes $^{69}\mathrm{Ga}$, $^{71}\mathrm{Ga}$, and $^{75}\mathrm{As}$ \cite{paget:1977a}.  

In ref. \cite{burkard:1999a}, the spin-flip probability was calculated perturbatively in the presence of a polarized nuclear spin system or applied magnetic field.  For a nuclear spin system composed of $N$ nuclear spins, of polarization $p\gg1/\sqrt{N}$, it was shown that this probability is suppressed by the factor $1/p^2 N$ in zero magnetic field.  The perturbative treatment, however, shows secular divergences and cannot be trusted, in general \cite{khaetskii:2002a}.  Subsequently, an exact solution has been found for the dynamics under $H_\mathrm{hf}$ when the nuclear spin system is fully polarized ($p=1$) \cite{khaetskii:2002a,khaetskii:2003a}.  The exact solution shows that the electron spin (longitudinal or transverse components) decay by only a fraction $1/N$ of the initial value due to the position-dependent coupling constants $A_i$, on a time scale $t\sim \hbar N/A$ ($t\simeq 1\,\mu \mathrm{s}$ for a GaAs quantum dot containing $N\simeq 10^5$ nuclei). Unlike the usual exponential decay that is seen in Markovian systems, this decay was found to be described by a power law $\sim1/t^{3/2}$ at long times, which suggests that the dynamics were non-Markovian (history dependent).  For an unpolarized nuclear spin system, treating the nuclei as a classical unpolarized fluctuating field which can be ensemble-averaged leads to a decay time $\tau_c(0) =\hbar\sqrt{N}/A$ ($\tau_c(0)\simeq 1-10\,\mathrm{ns}$ in a GaAs quantum dot containing $N=10^5-10^6$ nuclei \cite{khaetskii:2002a,merkulov:2002a}). Subsequently, it has been shown that the quantum nature of the nuclear spin distribution with polarization $p$ can lead to decay of the transverse electron spin in a high magnetic field on a time scale $\tau_c(p) =\hbar\sqrt{N}/A\sqrt{1-p^2}$, even without an ensemble average over initial conditions \cite{coish:2004a}. This transverse decay is, however, reversible with a spin echo pulse and can be removed by performing a projective (von Neumann) measurement on the nuclear Overhauser field \cite{coish:2004a}, which might be implemented experimentally using phase estimation \cite{giedke:2005a} or spin-selective Raman transitions \cite{stepanenko:2005a}.   Even if the nuclear system is perfectly prepared through measurement, or if an echo is performed, there is a non-reversible part of the electron spin decay due to flip-flop terms in the Hamiltonian ($V_\mathrm{ff}=(h_+S_-+h_-S_+)/2$), which become important in a weak magnetic field, where $g\mu_B B \lesssim A/\sqrt{N}$.  The effect of these terms has been evaluated to leading order for the longitudinal and transverse electron spin components, confirming the power-law decay of ref. \cite{khaetskii:2002a} for a nuclear spin system of reduced polarization $p<1$ \cite{coish:2004a}.  

Further studies of electron-spin decay due to the hyperfine interaction in quantum dots include numerical exact-diagonalization studies for spins without \cite{schliemann:2002a} and with \cite{shenvi:2005a} spin echo, lower bounds on the spin $T_2$ time in the presence of a magnetic field \cite{shenvi:2005b}, decay of the Hahn spin-echo envelope due to hyperfine in combination with nuclear dipolar interactions \cite{desousa:2003a,desousa:2003b, witzel:2005a, yao:2005a,yao:2006a}, the application of Carr-Purcell-Meiboom-Gill (CPMG) spin-echo pulses to extend the electron spin decay time \cite{coish:2004a, desousa:2005a}, decay through hyperfine interaction and phonon-assisted tunneling \cite{abalmassov:2004a}, nuclear-spin entanglement due to electron spin transport current and the hyperfine interaction \cite{eto:2004a}, slow oscillations of transport current through a double quantum dot due to the hyperfine interaction \cite{erlingsson:2005a} (which may be related to recent exciting experiments \cite{ono:2004a,koppens:2005a}), and semiclassical spin dynamics \cite{erlingsson:2004a,yuzbashyan:2005a}.  

Several experiments have now been performed on single electrons confined to ensembles of dots \cite{dzhioev:2002a} and single quantum dots \cite{dutt:2005a,bracker:2005a}.  These experiments have confirmed the ensemble-averaged transverse electron spin decay time $\tau_c \simeq \hbar\sqrt{N}/A\simeq1-10\,\mathrm{ns}$.  While these most recent measurements are done on individual quantum dots, the measurement times are still much longer than the typical evolution time of the nuclear spin system and therefore represent ensemble-averaged values.  If the measurement times of these experiments could be reduced further, they may sample single nuclear fields and thus show longer decay times.  This idea is connected to a quantum measurement (state narrowing) proposal that we describe further in the next section.  

\subsection{Two-spin decoherence, driven Rabi oscillations, and state narrowing}
\label{sec:HFTwoSpin}

Recent experiments have measured energy relaxation \cite{johnson:2005a} and dephasing \cite{petta:2005a,laird:2005a} for two-electron spin states in double quantum dots. The decay of electron spin states seen in these experiments can be understood in the presence of a moderate magnetic field using a simple two-level effective Hamiltonian.  In a large magnetic field such that $g\mu_\mathrm{B}B\gg A/\sqrt{N}$, flip-flop terms in the Hamiltonian $V_\mathrm{ff}$ that do not conserve $S^z$ can be neglected, and the Hamiltonian becomes block-diagonal with blocks labeld by $m_z$, the eigenvalues of $S^z$.  The evolution of a two-electron system in the subspace of $m_z=0$ spin triplet $\ket{T_0}=\left(\ket{\uparrow\downarrow}+\ket{\downarrow\uparrow}\right)/\sqrt{2}$ and spin singlet $\ket{S}=\left(\ket{\uparrow\downarrow}-\ket{\downarrow\uparrow}\right)/\sqrt{2}$ states is given by the effective Hamiltonian \cite{coish:2005a}
\begin{equation}
H_\mathrm{dd}=\frac{J}{2}\mathbf{S}\cdot\mathbf{S}+\delta h^z \delta S^z =\left(\begin{array}{cc} J & \delta h^z\\ \delta h^z & 0 \end{array}\right), 
\label{eq:HFDDHamiltonian}
\end{equation}
where $J$ is the exchange coupling, $\mathbf{S}=\mathbf{S}_1+\mathbf{S}_2$ is the total spin operator,  $\delta S^z=S^z_1-S^z_2$ is the spin-difference operator for the electron spins on dots 1 and 2, and $\delta h^z =(h^z_1-h^z_2)/2$ is the nuclear spin difference operator for nuclear fields on dots 1 and 2.  In recent experiments \cite{petta:2005a,laird:2005a}, a singlet intial state was prepared $\ket{\psi(0)}=\ket{S}$ and allowed to evolve under the action of the nuclear spins in the presence of a large magnetic field, as described by the effective Hamiltonian given in Eq. (\ref{eq:HFDDHamiltonian}). This Hamiltonian describes the competition of two effects; a large exchange $J$ tends to preserve the singlet nature of the initial state, but a finite difference in the hyperfine fields $\delta h^z\ne0$ will tend to cause precession between singlet and triplet states $\ket{S}\to\ket{T_0}\to\ket{S}\ldots$.  In these experiments, the measured quantity is the probability to remain in the singlet state at a later time, assuming the system begins in the singlet state at time $t=0$.  This probability is given by the correlator\footnote{The correlator $C_{\mathrm{SS}}(t)$ is related to $C_\mathrm{T_0}(t)$ (calculated in Ref. \cite{coish:2005a}) through $C_{\mathrm{SS}}(t)=1-C_\mathrm{T_0}(t)$} (setting $\hbar=0$)\cite{coish:2005a,klauser:2006a}
\begin{equation}
C_{\mathrm{SS}}(t)=\sum_n\rho(n)\left|\bra{n}\otimes\bra{S}e^{-iH_\mathrm{dd}t}\ket{S}\otimes\ket{n}\right|^2,
\label{eq:CSSValue}
\end{equation}   
where $\rho(n)$ gives the diagonal elements of the nuclear spin density matrix in the basis of $\delta h^z$ eigenstates: $\delta h^z\ket{n}=\delta h^z_n\ket{n}$.  For a sufficiently randomized system of $N\gg1$ nuclear spins, $\rho(n)$ describes a Gaussian distribution of $\delta h^z$ eigenvalues \cite{coish:2004a} with zero mean and variance $\sigma^2$.  Neglecting possible Poincar{\'e} recurrence, the sum in Eq. (\ref{eq:CSSValue}) can be converted to an integral, which can be evaluated in several interesting limits.  In particular, there is an abrupt crossover in the long-time dynamics of $C_\mathrm{SS}(t)$ from a rapid Gaussian to a slow power-law decay when the exchange becomes nonzero \cite{coish:2005a}: 
\begin{eqnarray}
C_\mathrm{SS}(t) & = & \frac{1}{2}e^{-2(\sigma t)^2}, J = 0, \label{eq:CSSJzero}\\
C_\mathrm{SS}(t) & \simeq & \frac{1}{2}-2\left(\frac{\sigma}{J}\right)^2+2\left(\frac{\sigma}{J}\right)^2\frac{\cos\left[Jt+\frac{3}{2}\arctan\left(\frac{4\sigma^2 t}{J}\right)\right]}{\left[1+\left(\frac{4\sigma^2 t}{J}\right)^2\right]^{3/4}}, J \gg \sigma.
\label{eq:CSSJlarge}
\end{eqnarray}
The decay time of $C_\mathrm{SS}(t)$ is given by $\tau_\mathrm{c,ST} \sim 1/\sigma\sim\sqrt{N}/A$ in the case of $J=0$ and by $\tau_\mathrm{c,ST} \sim J/\sigma^2$ for $J\gg\sigma$.  Due to the power-law decay $\left(C_\mathrm{SS}(t)\sim 1/t^{3/2}\right)$ at long times in Eq. (\ref{eq:CSSJlarge}), coherent oscillations in this correlator persist for a very long time, as has been observed in experiment \cite{petta:2005a,laird:2005a}.  Moreover, in the presence of exchange, the correlator acquires a universal (parameter independent) phase shift of $3\pi/4$ at long times. Other correlators corresponding to other initial and final states have also been calculated, and lead to power-law decay with a time dependence $\sim 1/t^{1/2}$ or $\sim 1/t^{3/2}$, and phase shifts of $\pi/4$ or $3\pi/4$, depending on the correlator \cite{klauser:2006a}.   

If the exchange coupling is modulated $J=J(t)$ with a frequency that matches the nuclear field inhomogeneity $\omega = \delta h^z_n$ (where $\delta h^z_n$ is an eigenvalue of the operator $\delta h^z$ corresponding to the eigenstate $\ket{n}$), the Hamiltonian in Eq. (\ref{eq:HFDDHamiltonian}) then describes resonant driven Rabi oscillations between the spin states $\ket{\uparrow\downarrow},\ket{\downarrow\uparrow}$.  Tuning the driving frequency $\omega$ through resonance to locate the point $\omega=\delta h^z_n$ thus determines the value of the Overhauser field inhomogeneity $\delta h^z$ and projects the state of the nuclear system into an eigenstate $\ket{n}$.  In this case, the decay of the correlator $C_\mathrm{SS}(t)$ can be removed, except for corrections to the effective Hamiltonian.  Such a measurement procedure has been described in detail in Ref. \cite{klauser:2006a}.

In very recent experiments \cite{koppens:2006a}, driven Rabi oscillations of a single electron spin have been generated in quantum dots using the more-conventional \emph{magnetic field} modulation (proposed in Refs. \cite{burkard:1999a,engel:2001a}, as aposed to exchange modulation, described above).  These coherent Rabi oscillations decay due to hyperfine coupling with the nuclear spin system according to a slow power-law, as in the case without driving, described in Refs. \cite{coish:2005a,klauser:2006a}, so coherent oscillations can be observed up to a time scale of $\sim\mu\mathrm{s}$ in spite of an inital partial decay that occurs on a time scale of $\tau_c=1-10\,\mathrm{ns}$\cite{koppens:2006b}.  These experiments are crucial for the realization of quantum-dot quantum computing since they provide a means to perform arbitrary single-spin (one-qubit) rotations on quantum-dot spins, one of the five DiVincenzo criteria. 

\section{Summary and conclusions}
\label{sec:Conclusions}

To conclude the review, we now give a brief discussion of relaxation and decoherence times (Secs. \ref{sec:SpinOrbit} and \ref{sec:Hyperfine}) as they apply directly to the proposal for Bell-state parity measurements (Sec. \ref{sec:BellState}).  Most relevant time scales have only very recently been measured in experiment.  Some of these times are presented in Table \ref{tab:DecayTimes}, below.   
\begin{table}[tbh]
  \caption{Experimentally determined energy relaxation and decoherence times that are relevant for the measurement-based quantum computing proposal described in section \ref{sec:BellState}\label{tab:DecayTimes}}
  \begin{tabular}{lll} \hline\hline
 & Decay type & Measured decay time(s)\\ \hline\hline

\begin{minipage}{0.05\textwidth}
\vspace{0.1 cm}
(i)
\end{minipage}
&
\begin{minipage}{0.3\textwidth}
\vspace{0.1 cm}
Energy relaxation time $T_1$ between Zeeman-split sublevels 
\end{minipage}
& \begin{minipage}{0.55\textwidth}
\vspace{0.1 cm}$T_1>50 \mu\mathrm{s}$\cite{hanson:2003a} (lateral GaAs dot),\\
$T_1=0.85\,\mathrm{ms}$ at $B=8\,\mathrm{T}$\cite{elzerman:2004a} (lateral GaAs dot),\\
$T_1=20\,\mathrm{ms}$ at $B=4\,\mathrm{T}$\cite{kroutvar:2004a} (self-assembled Ga(In)As dot),\\
$T_1=170\,\mathrm{ms}$ at $B=1.75\,\mathrm{T}$\cite{amasha:2006a} (lateral GaAs dot).
\end{minipage}\\ \hline

\begin{minipage}{0.05\textwidth}
\vspace{0.1 cm}
(ii)
\end{minipage}
&
\begin{minipage}{0.3\textwidth}
\vspace{0.1 cm}Single-spin coherence time $\tau_c$ due to the hyperfine interaction 
\end{minipage}
& \begin{minipage}{0.55\textwidth}
\vspace{0.1 cm}$\tau_c=1-10\mathrm{ns}$\cite{bracker:2005a,braun:2005a,dutt:2005a}
\end{minipage}\\ \hline

\begin{minipage}{0.05\textwidth}
\vspace{0.1 cm}
(iii)
\end{minipage}
&
\begin{minipage}{0.3\textwidth}
\vspace{0.1 cm}Singlet-triplet energy relaxation time $T_\mathrm{1,ST}$ 
\end{minipage}
& \begin{minipage}{0.55\textwidth}\vspace{0.1 cm}
$T_\mathrm{1,ST}\simeq 10-500\,\mu\mathrm{s}$\cite{fujisawa:2002a,sasaki:2005a} (vertical single dot),\\
$T_\mathrm{1,ST}\simeq 0.2-2.5\,\mathrm{ms}$\cite{hanson:2005a} (lateral single dot),\\
$T_\mathrm{1,ST}\simeq 1\mu\mathrm{s}-10\,\mathrm{ms}$\cite{petta:2005b,johnson:2005a} (lateral double dot)
\end{minipage}\\ \hline

\begin{minipage}{0.05\textwidth}
\vspace{0.1 cm}
(iv)
\end{minipage}
&
\begin{minipage}{0.3\textwidth}\vspace{0.1 cm}
Singlet-triplet coherence time due to hyperfine $\tau_\mathrm{c,ST}$ 
\end{minipage}
& \begin{minipage}{0.55\textwidth}\vspace{0.1 cm}
$\tau_\mathrm{c,ST}\simeq 10\,\mathrm{ns}$\cite{petta:2005a} (free-induction decay time)\\
$\tau_\mathrm{c,ST,echo}\simeq 1.2\,\mu\mathrm{s}$\cite{petta:2005a} (spin-echo envelope decay time)\\
\end{minipage}\\ \hline

\begin{minipage}{0.05\textwidth}
\vspace{0.1 cm}
(v)
\end{minipage}
&
\begin{minipage}{0.3\textwidth}\vspace{0.1 cm}
Orbital inelastic relaxation time $T_\mathrm{1,orb}$ for a single electron charge state in a double quantum dot 
\end{minipage}
& \begin{minipage}{0.55\textwidth}\vspace{0.1 cm}
$T_\mathrm{1,orb}\simeq 16\,\mathrm{ns}$\footnote{We note that this measurement was performed with a splitting supplied by the inter-dot tunnel coupling $t_\mathrm{d}$: $\epsilon\simeq 2t_\mathrm{d}$.  For the proposal of Ref. \cite{engel:2005a}, the relevant time scale is expected to be $T_\mathrm{1,orb}\simeq \frac{U^2\tau_\phi}{2t_\mathrm{d}^2}\gtrsim 1\mu\mathrm{s}$ \cite{engel:2005a} (for $t_\mathrm{d}=10\,\mu eV$ and splitting supplied by the on-site repulsion $\epsilon=U=1\,\mathrm{m}eV$)} \cite{petta:2004b}  
\end{minipage}\\ \hline

\begin{minipage}{0.05\textwidth}
\vspace{0.1 cm}
(vi)
\end{minipage}
&
\begin{minipage}{0.3\textwidth}\vspace{0.1 cm}
Orbital dephasing time $\tau_\phi$ for a single electron charge state in a double quantum dot 
\end{minipage}
& \begin{minipage}{0.55\textwidth}\vspace{0.1 cm}
$\tau_\phi\simeq 1\,\mathrm{ns}$\cite{hayashi:2003a},\\
$\tau_\phi\simeq 400\,\mathrm{ps}$\cite{petta:2004b},\\
(see also the comprehensive review in Ref. \cite{fujisawa:2006a})
\end{minipage}\\
\hline\hline
  \end{tabular}
  \label{tab:spindecaytimes}
\end{table}
We give an analysis of Table 1 row-by-row, below:

(i): The single-spin relaxation processes represent the time scale for a spin flip event ($\ket{\uparrow}\to\ket{\downarrow}$) to occur.  At large magnetic fields, the spin-flip rate found experimentally is well-described by spin flips due to the spin-orbit interaction and coupling to lattice phonons. In this case theory predicts a dependence $1/T_1\propto B^5$ \cite{khaetskii:2001a,golovach:2004a}, which has been verified experimentally \cite{kroutvar:2004a}.  At weaker magnetic fields, another mechanism involving a spin-flip due to the hyperfine interaction and phonon emission may dominate the relaxation time since this mechanism leads to a weaker field dependence $1/T_1\propto B^3$\cite{erlingsson:2002a}.

(ii) and (iv): The coherence time for a single spin $\tau_\mathrm{c}$ is the lifetime of a coherent linear combination $a\ket{\uparrow}+b\ket{\downarrow}$, and the singlet-triplet decoherence time $\tau_\mathrm{c,ST}$ is the lifetime of the linear combination $a\ket{S}+b\ket{T_0}$. Dephasing due to the hyperfine interaction presents a significant challenge to performing single-qubit rotations, although there are known ways to deal with this problem in the presence of a moderate magnetic field.  In particular, increasing the nuclear spin polarization to $p\gg1/\sqrt{N}$ reduces the phase-space available for nuclear field fluctuations, extending the free-induction decay time to $\tau_\mathrm{c}(p)=\frac{\tau_\mathrm{c}(0)}{\sqrt{1+p^2}}$ \cite{coish:2004a} (polarization schemes have been suggested \cite{imamoglu:2003a} for such a purpose).  Alternatively, a projective measurement could be performed on the nuclear spin system to remove the effect of the hyperfine fluctuations \cite{coish:2004a,giedke:2005a,klauser:2006a,stepanenko:2005a}. Another possibility is that a spin-echo sequence could be performed to recover the spin coherence during operations (note that the spin-echo sequence applied in Ref. \cite{petta:2005a} extended the decay time by almost three orders of magnitude).  However, a spin echo can only preserve a qubit for information storage.  Single-spin operations must still be performed on the time scale of free-induction decay.  

(iii): The singlet-triplet energy relaxation time is the timescale for the inelastic transition from triplet to singlet $\ket{T}\to\ket{S}$. This time can be made extremely long (up to 10 ms has been measured in a gated double dot).  These relaxation times are dominated by hyperfine interaction and phonon emission at low magnetic fields, as predicted in Ref. \cite{erlingsson:2001a} and confirmed experiementally in Ref. \cite{johnson:2005a}.

(v) and (vi): The orbital dephasing time $\tau_\phi$ is the lifetime of a coherent superposition of charge states $a\ket{L}+b\ket{R}$, where a charge is in a superposition of being on the right and left quantum dots, whereas the orbital relaxation time $T_\mathrm{1,orb}$ is the timescale for an inelastic transition between orbital levels.  Short orbital dephasing times $\tau_\phi$ compared to the measurement time are assumed in the analysis of Ref. \cite{engel:2005a} for the measurement procedure described in Fig. 1(c).  Thus, the readout time must be longer than the typical orbital dephasing time $\tau_\phi\sim 1\,\mathrm{ns}$, but still much shorter than the orbital relaxation time $T_\mathrm{1,orb}$ and the typical spin decoherence time, described by rows (ii) and (iv).  Since the hyperfine coupling can be controlled, in principle, up to time scales of $\sim\mu\mathrm{s}$, the relevant parameter regime may be achievable.

The requirements for quantum information processing with quantum-dot spins are impressive.  However, in spite of the myriad sources of decoherence and many different suggestions for manipulation, coupling, and readout of quantum-dot electron spins, there is still a flood of new proof-of-principle experiments \cite{elzerman:2004a,petta:2005a,koppens:2005a,koppens:2006a} that demonstrate the basic elements of a quantum-dot quantum computer.  The challenge for the future now lies in integrating some of these elements in a way that is scalable, and operating these quantum gates at a rate where error-correction procedures can be effective.   

\begin{acknowledgement}
We thank M. Borhani, D. Klauser, and M. Trif for useful discussions. We acknowledge financial support from the Swiss NSF, the NCCR Nanoscience, EU NoE MAGMANet, DARPA, ARO, ONR, CNPq, and JST ICORP.
\end{acknowledgement}

\bibliographystyle{prsty_allnames}
\bibliography{pssb.200642348}
\end{document}